\documentclass[11pt]{revtex4}    
\usepackage{amssymb,epsf}
\usepackage{latexsym}
\begin{document}
\title{Viscous dark energy and generalized second law of thermodynamics}
\author{M. R. Setare $^{1,2}$\footnote{rezakord@ipm.ir} and  A. Sheykhi$^{2,3}$\footnote{sheykhi@mail.uk.ac.ir}}
\address{$^1$Department of Science, Payame Noor University, Bijar, Iran
\\
         $^2$Research Institute for Astronomy and Astrophysics of Maragha (RIAAM), Maragha,
         Iran\\
$^{3}$ Department of Physics, Shahid Bahonar University, P.O. Box
76175, Kerman, Iran }
\begin{abstract}
We examine the validity of the generalized second law of
thermodynamics in a non-flat universe in the presence of viscous
dark energy. At first we assume that the universe filled only with
viscous dark energy. Then, we extend our study to the case where
there is an interaction between viscous dark energy and
pressureless dark matter. We examine the time evolution of the
total entropy, including the entropy associated with the apparent
horizon and the entropy of the viscous dark energy inside the
apparent horizon.  Our study show that the generalized  second law
of thermodynamics is always protected in a universe filled with
interacting viscous dark energy and dark matter in a region
enclosed by the apparent horizon. Finally, we show that the the
generalized second law of thermodynamics is fulfilled for a
universe filled with interacting viscous dark energy and dark
matter in the sense that we take into account the Casimir effect.
\end{abstract}
\maketitle

\section{Introduction\label{Intro}}
One of the most important problems of modern cosmology is the
so-called dark energy (DE) puzzle. The type Ia supernova
observations suggest that the universe is dominated by DE with
negative pressure which provides the dynamical mechanism for the
accelerating expansion of the universe \cite{Rie}. This acceleration
implies that if Einstein's theory of gravity is reliable on
cosmological scales, then our universe is dominated by a mysterious
form of energy. This unknown energy component possesses some strange
features, for example it is not clustered on large length scales and
its pressure must be negative so that can drive the current
acceleration of the universe. Since the fundamental theory of nature
that could explain the microscopic physics of DE is unknown at
present, phenomenologists take delight in constructing various
models based on its macroscopic behavior. The dynamical nature of
dark energy, at least in an effective level, can originate from
various fields, such is a canonical scalar field (quintessence)
\cite{quint}, a phantom field, that is a scalar field with a
negative sign of the kinetic term \cite{phant}, or the combination
of quintessence and phantom in a unified model named quintom
\cite{quintom}.

The cosmological models with non-viscous cosmic fluid has been
studied widely in the literature. Early treatises on viscous
cosmology are given in \cite{Pad}. The viscous entropy production
in the early universe and viscous fluids on the Randall-Sundrum
branes have been studied respectively in \cite{Bre0}. A special
branch of viscous cosmology is to investigate how the bulk
viscosity can influence the future singularity, commonly called
the Big Rip, when the fluid is in the phantom state corresponding
to $w_D<-1$. A lot of works have been done in this direction
\cite{Bre1,Bre2, meng}. In particular, it was first pointed out in
\cite{Bre1} that the presence of a bulk viscosity proportional to
the Hubble expansion $H$ can cause the fluid to pass from the
quintessence region into the phantom region and thereby
inevitably lead to a future singularity.

Since the discovery of black hole thermodynamics in $1970$,
physicists have been speculated on the thermodynamics of the
cosmological models in an accelerated expanding universe
\cite{Huan,Pavon2,Cai2,Cai3,CaiKim,Fro,Wang,Cai4}. Related to the
present work, the first and the second laws of thermodynamics in a
flat universe were investigated for time independent and time
dependent EoS \cite{bw}. For the case of a constant EoS, the first
law is valid for the apparent horizon (Hubble horizon) and it does
not hold for the event horizon as system's IR cut-off. When the
EoS is assumed to be time dependent, using a holographic model of
dark energy in flat space, the same result is gained; the event
horizon, in contrast to the apparent horizon, does not satisfy the
first law. Also, while the event horizon does not respect the
second
law, it hold for the universe enclosed by the apparent horizon.\\
In this paper we study the validity of the generalized second law
of thermodynamics for a viscous dark energy in a universe
enveloped by the apparent horizon. Recently, it was shown that for
an accelerating universe the apparent horizon is a physical
boundary from the thermodynamical point of view
\cite{Jia,Shey1,Shey2,sheywang}. In particular, it was argued that
for an accelerating universe inside the event horizon the
generalized second law does not satisfy, while the accelerating
universe enveloped by the apparent horizon satisfies the
generalized second law  of thermodynamics \cite{Jia}. Therefore,
the event horizon in an accelerating universe might not be a
physical boundary from the thermodynamical point of view. Then we
extend our study to the case where there is an interaction between
viscous dark energy and pressureless dark matter.  Most
discussions on dark energy rely on the assumption that it evolves
independently of dark matter. Given the unknown nature of both
dark energy and dark matter there is nothing in principle against
their mutual interaction and it seems very special that these two
major components in the universe are entirely independent. Indeed,
this possibility has received a lot of attention recently
\cite{Ame,Zim,Seta1,wang1} and in particular, it has been shown
that the coupling can alleviate the coincidence problem
\cite{Pav1}.

This paper is organized as follows. In section \ref{GSL}, we
examine the generalized second law of thermodynamics in a universe
filled only with viscous dark energy. In section \ref{GSLL}, we
extend our study to the case where there is an interaction term
between viscous dark energy and pressureless dark matter. In
section \ref{Casimir}, we study the Casimir effect in viscous dark
energy. The last section is devoted to conclusions.
\section{GSL and viscous dark energy \label{GSL}}
We start from a homogenous and isotropic
Friedmann-Robertson-Walker (FRW) universe which is described by
the line element
\begin{equation}
ds^2={h}_{\mu \nu}dx^{\mu}
dx^{\nu}+\tilde{r}^2(d\theta^2+\sin^2\theta d\phi^2),
\end{equation}
where $\tilde{r}=a(t)r$, $x^0=t, x^1=r$, the two dimensional
metric $h_{\mu \nu}$=diag $(-1, a^2/(1-kr^2))$. Here $k$ denotes
the curvature of space with $k = 0, 1, -1$ corresponding to open,
flat, and closed universes, respectively. A closed universe with a
small positive curvature ($\Omega_k\simeq0.01$) is compatible with
observations \cite{spe}. The dynamical apparent horizon, a
marginally trapped surface with vanishing expansion, is determined
by the relation $h^{\mu \nu}\partial_{\mu}\tilde
{r}\partial_{\nu}\tilde {r}=0$, which implies that the vector
$\nabla \tilde {r}$ is null on the apparent horizon surface. The
apparent horizon was argued as a causal horizon for a dynamical
spacetime and is associated with gravitational entropy and surface
gravity \cite{Hay2,Bak}. For the FRW universe the apparent horizon
radius reads
\begin{equation}
\label{radius}
 \tilde{r}_A=\frac{1}{\sqrt{H^2+k/a^2}}.
\end{equation}
The Friedmann equation for a non-flat universe filled with viscous
dark energy takes the form (we neglect the dark matter)
\begin{eqnarray}\label{Fried}
H^2+\frac{k}{a^2}=\frac{8\pi G}{3} \rho_D ,
\end{eqnarray}
where $\rho_D$ is the energy density of dark energy inside
apparent horizon. In an isotropic and homogeneous FRW universe,
the dissipative effects arise due to the presence of bulk
viscosity in cosmic fluids. The theory of bulk viscosity was
initially investigated by Eckart \cite{Eck} and later on pursued
by Landau and Lifshitz \cite{Lan}. Dark energy with bulk viscosity
has a peculiar property to cause accelerated expansion of phantom
type in the late evolution of the universe \cite{Bre1,Bre2}. It
can also alleviate several cosmological puzzles like age problem,
coincidence problem  and phantom crossing. The energy-momentum
tensor of the viscous fluid is
\begin{equation}\label{T}
T_{\mu\nu}=\rho_Du_{\mu}u_{\nu}+\tilde{p}_D(g_{\mu\nu}+u_{\mu}u_{\nu}),
\end{equation}
where $u_{\mu}$ is the four-velocity vector and
\begin{equation}\label{cons}
\tilde{p}_D={p}_D-3H\xi,
\end{equation}
is the effective pressure of dark energy and $\xi$ is the bulk
viscosity coefficient. We require $\xi>0$ to get positive entropy
production in conformity with second law of thermodynamics
\cite{Z}. The energy conservation equation is
\begin{eqnarray}
\dot{\rho}_D+3H(\rho_D+\tilde{p}_D)=0,\label{consq1}
\end{eqnarray}
which can be written
\begin{eqnarray}
\dot{\rho}_D+3H\rho_D(1+w_D)=9H^2\xi,\label{consq2}
\end{eqnarray}
where $w_{D}=p_D/\rho_D$ is the equation of state parameter of
viscous dark energy. In terms of the apparent horizon radius, we
can rewrite the Friedmann equation as
\begin{equation}
\label{Frid2}
 \frac{1}{\tilde {r}_{A}^2}=\frac{8\pi G}{3}\rho_D.
 \end{equation}
The associated surface gravity on the apparent horizon can be
defined as
\begin{equation}
\label{surgra}\label{kappa}
 \kappa =\frac{1}{\sqrt{-h}}\partial_{a}\left(\sqrt{-h}h^{ab}\partial_{ab}\tilde
 {r}\right).
\end{equation}
Then one can easily show that the surface gravity at the apparent
horizon of FRW universe can be written as
\begin{equation}\label{surgrav}
\kappa=-\frac{1}{\tilde r_A}\left(1-\frac{\dot {\tilde
r}_A}{2H\tilde r_A}\right).
\end{equation}
The associated temperature on the apparent horizon can be defined
as
\begin{equation}\label{Therm}
T_{h} =\frac{|\kappa|}{2\pi}=\frac{1}{2\pi \tilde
r_A}\left(1-\frac{\dot {\tilde r}_A}{2H\tilde r_A}\right).
\end{equation}
where $\frac{\dot{\tilde r}_A}{2H\tilde r_A}<1$ ensures that the
temperature is positive. Recently the connection between
temperature on the apparent horizon and the Hawking radiation has
been observed in \cite{cao}. Hawking radiation is an important
quantum phenomenon of black hole, which is closely related to the
existence of event horizon of black hole.  The cosmological event
horizon of de Sitter space has the Hawking radiation with thermal
spectrum as well. Using the tunneling approach proposed by Parikh
and Wilczek, the authors of \cite{cao} showed that there is indeed
a Hawking radiation with a finite temperature, for locally defined
apparent horizon of the FRW universe with any spatial curvature.
This gives more solid physical implication of the temperature
associated with the apparent horizon. The entropy associated to
the apparent horizon is
\begin{eqnarray}\label{ent1}
S_{h} =\frac{A}{4G}=\frac{\pi \tilde{r}_{A}^2 }{G}.
\end{eqnarray}
where $A=4\pi\tilde{r}_{A}^{2}$ is the area of the apparent
horizon. Differentiating Eq. (\ref{Frid2}) with respect to the
cosmic time and using Eq. (\ref{consq2}) we get
\begin{equation} \label{dotrv1}
\dot{\tilde{r}}_{A}=4\pi G H {\tilde{r}_{A}^3}
\left[\rho_D(1+w_D)-3H\xi\right].
\end{equation}
Let us now turn to find out $T_{h} \dot{S_{h}}$:
\begin{equation}\label{TSh}
T_{h} \dot{S_{h}} =\frac{1}{2\pi \tilde r_A}\left(1-\frac{\dot
{\tilde r}_A}{2H\tilde r_A}\right)\frac{d}{dt} \left(\frac{\pi
\tilde{r}_{A}^2 }{G}\right).
\end{equation}
After some simplification and using Eq. (\ref{dotrv1}) we get
\begin{equation}\label{TShv1}
T_{h} \dot{S_{h}} =4\pi H {\tilde{r}_{A}^3}
\left[\rho_D(1+w_D)-3H\xi\right]\left(1-\frac{\dot {\tilde
r}_A}{2H\tilde r_A}\right).
\end{equation}
As we argued above the term $\left(1-\frac{\dot {\tilde
r}_A}{2H\tilde r_A}\right)$ is positive to ensure $T_{h}>0$,
however, in an accelerating universe the equation of state
parameter of dark energy may satisfy $w_D<-1+3H\xi/\rho_D$. This
indicates that the second law of thermodynamics,
$\dot{S_{h}}\geq0$, does not hold on the apparent horizon. Then
the question arises, ``will the generalized second law of
thermodynamics, $\dot{S_{h}}+\dot{S_{D}}\geq0$, can be satisfied
in a region enclosed by the apparent horizon?'' The entropy of the
viscous dark energy inside the apparent horizon, $S_{D}$, can be
related to its energy $E_D=\rho_D V$ and its pressure
$\tilde{p}_D$ by the Gibbs equation \cite{Pavon2}
\begin{equation}\label{Gibv1}
T_D dS_D=d(\rho_D V)+\tilde{p}_DdV=V d\rho_D+(\rho_D+p_D-3H\xi)dV,
\end{equation}
where $T_D$ and is the temperature of the viscous dark energy and
$V=\frac{4\pi}{3}\tilde{r}_{A}^{3}$ is the volume enveloped by the
apparent horizon. We also limit ourselves to the assumption that
the thermal system bounded by the apparent horizon remains in
equilibrium so that the temperature of the system must be uniform
and the same as the temperature of its boundary. This requires
that the temperature $T_D$ of the viscous dark energy inside the
apparent horizon should be in equilibrium with the temperature
$T_h$ associated with the apparent horizon, so we have $T_D=T_h$.
This expression holds in the local equilibrium hypothesis. If the
temperature of the fluid differs much from that of the horizon,
there will be spontaneous heat flow between the horizon and the
fluid and the local equilibrium hypothesis will no longer hold.
This is also at variance with the FRW geometry. In general, when
we consider the thermal equilibrium state of the universe, the
temperature of the universe is associated with the apparent
horizon. Therefore from the Gibbs equation (\ref{Gibv1}) we can
obtain
\begin{equation}\label{TSmv1}
T_{h} \dot{S_{D}} =4\pi {\tilde{r}_{A}^2}
\left[\rho_D(1+w_D)-3H\xi\right]\dot{\tilde{r}}_{A}-4\pi H
{\tilde{r}_{A}^3} \left[\rho_D(1+w_D)-3H\xi\right].
\end{equation}
To check the generalized second law of thermodynamics, we have to
examine the evolution of the total entropy $S_h + S_D$. Adding
equations (\ref{TShv1}) and (\ref{TSmv1}),  we get
\begin{equation}\label{GSLv1}
T_{h}( \dot{S}_{h}+\dot{S}_D)=2\pi {\tilde{r}_{A}^2}
\left[\rho_D(1+w_D)-3H\xi\right]\dot{\tilde{r}}_A=\frac{A}{2}\left[\rho_D(1+w_D)-3H\xi\right]
\dot {\tilde r}_A.
\end{equation}
where $A>0$ is the area of apparent horizon. Finally, substituting
$\dot {\tilde r}_A$ from Eq. (\ref{dotrv1}) into (\ref{GSLv1}) we
reach
\begin{equation}\label{GSLv12}
T_{h}( \dot{S}_{h}+\dot{S}_D)=2\pi G A H {\tilde r_A}^{3}
\left[\rho_D(1+w_D)-3H\xi\right]^2.
\end{equation}
The right hand side of the above equation cannot be negative
throughout the history of the universe, which means that $
\dot{S_{h}}+\dot{S_{D}}\geq0$ always holds. This indicates that
for a universe with spacial curvature filled with viscous dark
energy, the generalized second law of thermodynamics is fulfilled
in a region enclosed by the apparent horizon.

\section{GSL and interacting viscous dark energy with non-viscous dark matter \label{GSLL}}
In this section we extend our study to the case where there is an
interaction between viscous dark energy and pressureless dark
matter. In this case the Friedmann equation can be written as
\begin{eqnarray}\label{Fried}
H^2+\frac{k}{a^2}=\frac{8\pi G}{3} \left( \rho_m+\rho_D \right),
\end{eqnarray}
where $\rho_m$ and $\rho_D$ are the energy density of dark matter
and dark energy inside apparent horizon, respectively. Since we
consider the interaction between dark matter and dark energy,
$\rho_{m}$ and $\rho_{D}$ do not conserve separately, they must
rather enter the energy balances
\begin{eqnarray}
&&\dot{\rho}_m+3H\rho_m=Q, \label{consm}
\\&& \dot{\rho}_D+3H\rho_D(1+w_D)=9H^2\xi-Q.\label{consq}
\end{eqnarray}
where $Q=\Gamma \rho_D$ denotes the interaction between the dark
components. We also assume the interaction term is positive,
$Q>0$, which means that there is an energy transfer from the dark
energy to dark matter. In terms of the apparent horizon radius, we
can rewrite the Friedmann equation as
\begin{equation}
\label{Fri2}
 \frac{1}{\tilde {r}_{A}^2}=\frac{8\pi G}{3}\left(\rho_m+\rho_D\right).
 \end{equation}
Differentiating Eq. (\ref{Fri2}) with respect to the cosmic time
and using Eqs. (\ref{consm}) and (\ref{consq}) we get
\begin{equation} \label{dotr1}
\dot{\tilde{r}}_{A}=4\pi G H {\tilde{r}_{A}^3}
\left[\rho_D(1+u+w_D)-3H\xi\right].
\end{equation}
where $u =\rho_m/\rho_D$ is the ratio of energy densities. Next we
turn to calculate $T_{h} \dot{S_{h}}$. It is easy to show that
\begin{equation}\label{TSh1}
T_{h} \dot{S_{h}} =4\pi H {\tilde{r}_{A}^3}
\left[\rho_D(1+u+w_D)-3H\xi\right]\left(1-\frac{\dot {\tilde
r}_A}{2H\tilde r_A}\right).
\end{equation}
Again in an accelerating universe the equation of state parameter
of dark energy may satisfy the condition $w_D<-1-u+3H\xi/\rho_D$.
This implies that the second law of thermodynamics,
$\dot{S_{h}}\geq0$, does not hold on the apparent horizon. Then we
examine the validity of the generalized second law,
$\dot{S_{h}}+\dot{S_{m}}+\dot{S_{D}}\geq0$. The entropy of the
viscous dark energy plus dark matter inside the apparent horizon,
$S=S_{m}+S_{D}$, can be related to the total energy
$E=(\rho_m+\rho_D) V$ and pressure $\tilde{p}_D$ in the horizon by
the Gibbs equation
\begin{equation}\label{Gib1}
T dS=d[(\rho_m+\rho_D) V]+\tilde{p}_DdV=V(
d\rho_m+d\rho_D)+\left[\rho_D(1+u+w_D)-3H\xi\right]dV,
\end{equation}
where $T=T_{m}=T_D$ and $S=S_{m}+S_D$ are the temperature and the
total entropy of the energy and matter content inside the horizon,
respectively. Here we assumed that the temperature of both dark
components are equal, due to their mutual interaction. We also
assume the local equilibrium hypothesis holds, so $T=T_h$.
Therefore from the Gibbs equation (\ref{Gib1}) we obtain
\begin{equation}\label{TSm1}
T_{h} (\dot{S_{m}}+\dot{S_{D}}) =4\pi {\tilde{r}_{A}^2}
\left[\rho_D(1+u+w_D)-3H\xi\right]\dot{\tilde{r}}_{A}-4\pi H
{\tilde{r}_{A}^3} \left[\rho_D(1+u+w_D)-3H\xi\right].
\end{equation}
To check the generalized second law of thermodynamics, we have to
examine the evolution of the total entropy $S_h + S_m+S_D$. Adding
equations (\ref{TSh1}) and (\ref{TSm1}),  we get
\begin{equation}\label{GSL1}
T_{h}( \dot{S}_{h}+\dot{S}_{m}+\dot{S}_D)=2\pi {\tilde{r}_{A}^2}
\left[\rho_D(1+u+w_D)-3H\xi\right]\dot{\tilde{r}}_A=\frac{A}{2}\left[\rho_D(1+u+w_D)-3H\xi\right]
\dot {\tilde r}_A.
\end{equation}
Substituting $\dot {\tilde r}_A$ from Eq. (\ref{dotr1}) into
(\ref{GSL1}) we get
\begin{equation}\label{GSL2}
T_{h}( \dot{S}_{h}+\dot{S}_{m}+\dot{S}_D)=2\pi G A H {\tilde
r_A}^{3} \left[\rho_D(1+u+w_D)-3H\xi\right]^2,
\end{equation}
which cannot be negative throughout the history of the universe
and hence the general second law of thermodynamics,
$\dot{S_{h}}+\dot{S_{m}}+\dot{S_{D}}\geq0$, is always protected
for a universe filled with interacting viscous dark energy and
dark matter in a region enclosed by the apparent horizon. To see
the effect on the generalized second law of thermodynamics derived
from the interaction $Q$, one can consider the $Q=0$ in Eqs. (\ref
{consm}), (\ref{consq}). After this substituation, our result
(\ref{GSL2}) do not change, so we conclude that the interaction
term does not affect on the generalized second law of
thermodynamics.

\section{Casimir effects in viscous cosmology \label{Casimir}}
 In this section
we would like to examine the GSL of thermodynamics for an
interacting viscous dark energy in the sense that we take into
account the Casimir effect. A natural way of dealing with the
Casimir effect in a non-flat universe is to relate it to the
apparent horizon radius $\tilde{r}_{A}=1/\sqrt{H^2+k/a^2}$. It means
effectively that we should put the Casimir energy $E_c$ inversely
proportional to the apparent horizon radius. This is consistent with
the basic property of the Casimir energy, which states that it is a
measure of the stress in the region interior to the ``shell" as
compared with the unstressed region on the outside. The effect is
evidently largest in the beginning of the universe's evolution, when
$\tilde{r}_{A}$ is small. At late times, when
$\tilde{r}_{A}\rightarrow\infty$, the Casimir influence should be
expected to fade away. Therefore, we assume the Casimir energy can
be written as
\begin{eqnarray}\label{Ec}
E_c=\frac{c}{\tilde{r}_{A}},
\end{eqnarray}
where $c$ is a constant. We also assume that $c$ is small compared
with unity. This is physically reasonable, in view of the
conventional feebleness of the Casimir force. The Casimir pressure
corresponding to energy (\ref{Ec}) is
\begin{eqnarray}\label{pc}
p_c=\frac{-1}{4\pi\tilde{r}_{A}^2}\frac{\partial {E_c}}{\partial
{\tilde{r}_{A}}}=\frac{c}{4\pi \tilde{r}_{A}^4 }.
\end{eqnarray}
Thus the Casimir energy evolves as $\rho_c\propto
\tilde{r}_{A}^{-4}$. The continuity equation for the Casimir energy
takes the form
\begin{eqnarray}
\dot{\rho}_c+3H\rho_c(1+w_c)=0,\label{consc}
\end{eqnarray}
where $w_{c}=p_c/\rho_c$ is the equation of state parameter of
Casimir energy. Using Eq. (\ref{pc}) as well as relation
\begin{eqnarray}
\rho_c=\frac{E_c}{V}=\frac{3c}{4\pi \tilde{r}_{A}^4},
\end{eqnarray}
we have
\begin{eqnarray}
w_c=\frac{p_c}{\rho_c}=\frac{1}{3}.\label{wc}
\end{eqnarray}
The Friedmann equation now takes the form
\begin{eqnarray}\label{Fried}
H^2+\frac{k}{a^2}=\frac{8\pi G}{3} \left( \rho_m+\rho_D+
\rho_c\right),
\end{eqnarray}
which can be rewritten as
\begin{equation}
\label{Fri3}
 \frac{1}{\tilde {r}_{A}^2}=\frac{8\pi G}{3}\left(\rho_m+\rho_D+\rho_c\right).
 \end{equation}
Differentiating Eq. (\ref{Fri3}) with respect to the cosmic time and
using Eqs. (\ref{consm}), (\ref{consq}), (\ref{consc}) and
(\ref{wc}) we find
\begin{equation} \label{dotr2} \dot{\tilde{r}}_{A}=4\pi G H
{\tilde{r}_{A}^3} \left[\rho_D(1+u+\frac{4z}{3}+w_D)-3H\xi\right],
\end{equation}
where $z =\rho_c/\rho_D$. Next we calculate $T_{h} \dot{S_{h}}$. It
is a matter of calculation to show
\begin{equation}\label{TSh1c}
T_{h} \dot{S_{h}} =4\pi H {\tilde{r}_{A}^3}
\left[\rho_D(1+u+\frac{4z}{3}+w_D)-3H\xi\right]\left(1-\frac{\dot
{\tilde r}_A}{2H\tilde r_A}\right).
\end{equation}
From the Gibbs equation for the total energy content of the universe
we have
\begin{eqnarray}\label{Gib1c}
T_h dS&=&d[(\rho_m+\rho_D+\rho_c) V]+(\tilde{p}_D+p_c) dV \nonumber \\
&=&V(
d\rho_m+d\rho_D+d\rho_c)+\left[\rho_D(1+u+\frac{4z}{3}+w_D)-3H\xi\right]dV,
\end{eqnarray}
where $S=S_{m}+S_D+S_c$ and we have assumed that the temperature of
all the energy content are identical and equal with the apparent
horizon temperature $T_h$. Thus from Eq. (\ref{Gib1c}) we obtain
\begin{eqnarray}\label{TSc}
T_{h} (\dot{S_{m}}+\dot{S_{D}}+\dot{S_{c}}) &=&4\pi
{\tilde{r}_{A}^2}
\left[\rho_D(1+u+\frac{4z}{3}+w_D)-3H\xi\right]\dot{\tilde{r}}_{A}
\nonumber \\ && -4\pi H {\tilde{r}_{A}^3}
\left[\rho_D(1+u+\frac{4z}{3}+w_D)-3H\xi\right].
\end{eqnarray}
Now we are in a position to examine the GSL of thermodynamics.
Adding equations (\ref{TSh1c}) and (\ref{TSc}), we get
\begin{eqnarray}\label{GSL1c}
T_{h}( \dot{S}_{h}+\dot{S}_{m}+\dot{S}_D+\dot{S}_c)&=&2\pi
{\tilde{r}_{A}^2}
\left[\rho_D(1+u+\frac{4z}{3}+w_D)-3H\xi\right]\dot{\tilde{r}}_A
\nonumber \\
&=&\frac{A}{2}\left[\rho_D(1+u+\frac{4z}{3}+w_D)-3H\xi\right] \dot
{\tilde r}_A.
\end{eqnarray}
Substituting $\dot {\tilde r}_A$ from Eq. (\ref{dotr2}) into
(\ref{GSL1c}) we reach
\begin{equation}\label{GSL2c}
T_{h}( \dot{S}_{h}+\dot{S}_{m}+\dot{S}_D+\dot{S}_c)=2\pi G A H
{\tilde r_A}^{3} \left[\rho_D(1+u+\frac{4z}{3}+w_D)-3H\xi\right]^2.
\end{equation}
The right hand side of the above equation cannot be negative
throughout the history of the universe, which means that $
\dot{S_{h}}+\dot{S_{m}}+\dot{S}_D+\dot{S}_c\geq0$ always holds. This
indicates that the GSL of thermodynamics is fulfilled for a universe
filled with interacting viscous dark energy and dark matter in the
sense that we take into account the Casimir effect.

\section{Conclusions\label{Con}}
We have investigated the validity of the generalized second law of
thermodynamics in a non-flat universe with viscous dark energy. We
have examined the total entropy evolution with time, including the
derived apparent horizon entropy and the entropy of viscous dark
energy inside the apparent horizon. Then, we have extended our study
to the case where there is an interaction between viscous dark
energy and pressureless dark matter. We have shown that the
generalized second law of thermodynamics is always fulfilled for a
universe filled with interacting viscous dark energy and dark matter
in a region enclosed by the apparent horizon. We have also examined
the validity of the GSL of thermodynamics for an interacting viscous
dark energy in the sense that we take into account the Casimir
effect.

 \acknowledgments {This work has been supported
by Research Institute for Astronomy and Astrophysics of Maragha.}


\begin{thebibliography}{99}


\bibitem{Rie} A.G. Riess, et al., Astron. J.  116 (1998)
1009;\\
  S. Perlmutter, et al.,  Astrophys. J.  517 (1999) 565;\\
  S. Perlmutter, et al.,  Astrophys. J.  598 (2003) 102;\\
  P. de Bernardis, et al.,  Nature  404 (2000) 955.

\bibitem{quint}
B.~Ratra and P.~J.~E.~Peebles, Phys.\ Rev.\ D {\bf 37}, 3406 (1988);
C.~Wetterich, Nucl.\ Phys.\ B {\bf 302}, 668 (1988); A.~R.~Liddle
and R.~J.~Scherrer, Phys.\ Rev.\ D {\bf 59}, 023509 (1999);
I.~Zlatev, L.~M.~Wang and P.~J.~Steinhardt, Phys.\ Rev.\ Lett.\ {\bf
82}, 896 (1999).

\bibitem{phant} R. R. Caldwell, Phys.
Lett. B {\bf{545}}, 23 (2002); R.~R.~Caldwell, M.~Kamionkowski and
N.~N.~Weinberg, Phys. Rev. Lett. {\bf 91}, 071301 (2003); S. Nojiri
and S. D. Odintsov, Phys. Lett. B {\bf 562}, 147 (2003); V. K.
Onemli and R. P. Woodard, Phys.\ Rev.\ D {\bf 70}, 107301 (2004); M.
R. Setare, J. Sadeghi, A. R. Amani, Phys. Lett. B {\bf 666}, 288,
(2008);
  M.~R.~Setare and E.~N.~Saridakis,
  JCAP {\bf 0903}, 002 (2009).
\bibitem{quintom}
B.~Feng, X.~L.~Wang and X.~M.~Zhang, Phys.\ Lett.\  B {\bf 607}, 35
(2005);
Z. K. Guo, {\it{et al.}}, Phys. Lett. B {\bf 608}, 177 (2005); M.-Z
Li, B. Feng, X.-M Zhang, JCAP, 0512, 002 (2005); B. Feng, M. Li,
Y.-S. Piao and X. Zhang, Phys. Lett. B {\bf 634}, 101 (2006); M. R.
Setare, Phys. Lett. B {\bf 641}, 130 (2006); W. Zhao and Y. Zhang,
Phys. Rev. D {\bf73}, 123509 (2006);
 M. R.
Setare, J. Sadeghi, and A. R. Amani, Phys. Lett. B {\bf 660}, 299
(2008); J. Sadeghi, M. R. Setare, A. Banijamali and F. Milani, Phys.
Lett. B {\bf 662}, 92 (2008);
  M.~R.~Setare and E.~N.~Saridakis,
  Phys.\ Lett.\  B {\bf 668}, 177 (2008);
  M.~R.~Setare and E.~N.~Saridakis,
  JCAP {\bf 0809}, 026 (2008);
  M.~R.~Setare and E.~N.~Saridakis,
  Int.\ J.\ Mod.\ Phys.\  D {\bf 18}, 549 (2009).
\bibitem{Pad}T. Padmanabhan and S. M. Chitre, Phys. Lett. A 120, 433 (1987).
\bibitem{Bre0}I. Brevik and L. T. Heen, Astrophys. Space Sci. 219, 99 (1994);\\
 Brevik and A. Hallanger, Phys. Rev. D 69, 024009 (2004).
\bibitem{Bre1} I. Brevik and O. Gorbunova, Gen. Relativ. Gravit. 37, 2039
(2005).

\bibitem{Bre2} I. Brevik, O. Gorbunova and Y. A. Shaido, Int. J. Mod.
Phys. D 14, 1899 (2005);\\
I. Brevik and O. Gorbunova, Eur. Phys. J. C 56, 425 (2008);\\ I.
Brevik, Eur. Phys. J. C 56, 579 (2008).
\bibitem{meng}M. Cataldo, N. Cruz and S. Lepe, Phys. Lett. B  619, 5 (2005);
 M. Szydlowski and O. Hrycyna, Annals Phys. 322, 2745 (2007);
  X. H. Meng, J. Ren and M. G. Hu, Commun. Theor. Phys.  47, 379 (2007);
   X. H. Meng and X. Dou,
 arXiv:0910.2397 [astro-ph].
\bibitem{Huan}Q. Huang and M. Li, JCAP 0408, 013 (2004);\\
R. Brustein, Phys. Rev. Lett. 84, 2072 (2000);\\
P. F. Gonzalez-Diaz, hep-th/0411070;\\ P. C. W. Davies, Class.
Quant. Grav 5, 1349 (1988).

\bibitem{Pavon2} G. Izquierdo and D. Pavon, Phys.Lett. B 633 (2006)
420.

\bibitem{Cai2} M.~Akbar and R.~G.~Cai, Phys. Rev. D {\bf 75}, 084003 (2007).
  \bibitem{Cai3} R.~G.~Cai and L.~M.~Cao, Phys.Rev. D {\bf 75}, 064008
  (2007).

\bibitem{CaiKim} R. G. Cai and S. P. Kim, JHEP {\bf0502}, 050
(2005).

 \bibitem{Fro} A. V. Frolov and L. Kofman, JCAP {\bf 0305},
009 (2003);\\ U. K. Danielsson, Phys. Rev. D {\bf71}, 023516(2005)
;\\ R. Bousso, Phys. Rev. D {\bf71}, 064024 (2005);\\ G. Calcagni,
JHEP {\bf0509}, 060 (2005).
\bibitem{Wang} B. Wang, E.
Abdalla and R. K. Su, Phys.Lett. B {\bf503},  394 (2001);\\ B.
Wang, E. Abdalla and R. K. Su, Mod. Phys. Lett. A {\bf17},  23
(2002);\\ R.~G.~Cai and Y.~S.~Myung, Phys.\ Rev.\ D {\bf 67},
124021 (2003).
\bibitem{Cai4} R.~G.~Cai and L.~M.~Cao,
  Nucl. Phys. B {\bf785} (2007) 135.

\bibitem{bw} B. Wang, Y. Gong, E. Abdalla, gr-qc/0511051.
\bibitem{Jia} J. Zhou, B. Wang, Y. Gong, E. Abdalla, Phys.
Lett. B 652 (2007) 86.

 \bibitem{Shey1} A. Sheykhi, B. Wang and R. G. Cai, Nucl. Phys. B {\bf
779} (2007)1.
  \bibitem{Shey2} A. Sheykhi, B. Wang and R. G. Cai, Phys. Rev. D {\bf
76} (2007) 023515; A. Sheykhi, JCAP 05 (2009) 019.


\bibitem{sheywang} A. Sheykhi, B. Wang, Phys. Lett. B 678
(2009) 434;\\ A. Sheykhi, B. Wang,  Mod. Phys. Lett. A  Vol. 25,
No. 14 (2010) in press.



\bibitem{Ame} L. Amendola, Phys. Rev. D 60 (1999)  043501; \\ L. Amendola, Phys. Rev. D 62 (2000) 043511;
 \\ L. Amendola and C. Quercellini, Phys. Rev. D 68
(2003)  023514; \\ L. Amendola and D. Tocchini-Valentini, Phys. Rev.
D 64 (2001)  043509 ;\\ L. Amendola and D. T. Valentini, Phys. Rev.
D 66 (2002)  043528.


\bibitem{Zim} W. Zimdahl, D. Pavon, L.P. Chimento, Phys. Lett. B 521 (2001) 133;\\ W. Zimdahl and D. Pavon, Gen. Rel. Grav. 35
(2003) 413;\\ L. P. Chimento, A. S. Jakubi, D. Pavon and W. Zimdahl,
Phys. Rev. D 67 (2003)  083513.

\bibitem{Seta1}
M. R. Setare, Eur. Phys. J. C 50 (2007) 991;\\ M. R. Setare, JCAP
0701 (2007) 023;\\ M. R. Setare, Phys. Lett. B 654 (2007) 1;\\
M. R. Setare, Phys. Lett. B 642  (2006) 421.

\bibitem{wang1} B. Wang, Y. Gong and E. Abdalla, Phys. Lett. B 624
(2005) 141;\\ B. Wang, C. Y. Lin. D. Pavon and E. Abdalla, Phys.
Lett. B 662 (2008) 1.

\bibitem{Pav1} D. Pavon, W. Zimdahl, Phys. Lett. B 628 (2005) 206.

\bibitem{spe} D. N. Spergel, Astrophys. J. Suppl. 148 (2003) 175;\\
C. L. Bennett, et al.,  Astrophys. J. Suppl. 148 (2003) 1; \\ U.
Seljak, A. Slosar, P. McDonald, JCAP 0610 (2006) 014;\\ D. N.
Spergel, et al., Astrophys. J. Suppl. 170 (2007) 377.

\bibitem{Hay2} S.A. Hayward, S. Mukohyana, and M. C. Ashworth, Phys.
Lett.  A {\bf 256}, 347 (1999);\\ S. A. Hayward, Class. Quantum
Grav. {\bf 15}, 3147 (1998).
\bibitem{Bak} D. Bak and S. J. Rey, Class. Quantum Grav. {\bf17}, L83 (2000).


\bibitem{Eck} C. Eckart, Phys. Rev. 58 (1940) 919.

\bibitem{Lan} L.D. Landau and E.M. Lifshitz, Fluid Mechanics (Butterworth
Heineman, 1987)


\bibitem{Z} W. Zimdahl and D. Pavon, Phys. Rev. D 61 (2000)
108301.

\bibitem{cao}  R.G. Cai, L.M. Cao, Y.P. Hu, arXiv:0809.1554; \\
 R. Li, J. R. Ren, D. F. Shi, Phys. Lett. B {\bf670} (2009) 446.

\bibitem{Bre3} I. Brevik, O. Gorbunova, D. S. Gomez,
arXiv:0908.2882.

\end{thebibliography}
\end{document}